\documentclass[final]{ws-ijqi}

\usepackage{bbm,amsmath,amssymb,amsbsy,graphicx,subfigure,indentfirst}

\newcommand{\conc}{{\cal C}}

\newcommand{\n}[1]{\;\!\diagup \!\!\!\!\!#1}

\renewcommand{\det}{{\rm Det}\,}
\newcommand{\gr}[1]{\boldsymbol{#1}}
\newcommand{\be}{\begin{equation}}
\newcommand{\ee}{\end{equation}}

\newcommand{\bea}{\begin{eqnarray}}
\newcommand{\eea}{\end{eqnarray}}

\newcommand{\ket}[1]{|#1\rangle}

\newcommand{\ketbra}[2]{\vert #1 \rangle \! \langle #2 \vert}

\newcommand{\N}{{\cal N}}

\newcommand{\sig}{\gr{\sigma}}

\newcommand{\eq}[1]{Eq.~(\ref{#1})}
\newcommand{\ineq}[1]{Ineq.~(\ref{#1})}
\newcommand{\eg}{\emph{e.g.}~}
\newcommand{\ie}{\emph{i.e.}~}

\begin{document}

\markboth{G. Adesso \& F. Illuminati} {Entanglement Sharing: From
Qubits to Gaussian States}

\catchline{}{}{}{}{}

\title{ENTANGLEMENT SHARING: \\FROM QUBITS TO GAUSSIAN STATES}

\author{GERARDO ADESSO$^\ast$ and FABRIZIO ILLUMINATI$^\dag$}

\address{Dipartimento di Fisica ``E. R. Caianiello'',
Universit\`a di Salerno; \\CNR-Coherentia, Gruppo di Salerno; and
INFN Sezione di Napoli, \\Gruppo Collegato di Salerno, Via S.
Allende,
84081 Baronissi (SA), Italy.\\
$^\ast$gerardo@sa.infn.it \\$^\dag$illuminati@sa.infn.it}

\maketitle

\begin{history}
\received{June 24, 2005}
\end{history}

\begin{abstract}
It is a central trait of quantum information theory that there exist
limitations to the free sharing of quantum correlations among
multiple parties. Such {\em monogamy constraints} have been
introduced in a landmark paper by Coffman, Kundu and Wootters, who
derived a quantitative inequality expressing a trade-off between the
couplewise and the genuine tripartite entanglement for states of
three qubits. Since then, a lot of efforts have been devoted to the
investigation of distributed entanglement in multipartite quantum
systems. In these proceedings we report, in a unifying framework, a
bird's eye view of the most relevant results that have been
established so far on entanglement sharing. We will take off from
the domain of $N$ qubits, graze qudits, and finally land in the
almost unexplored territory of multimode Gaussian states of
continuous variable systems.
\end{abstract}




\section{Coffman-Kundu-Wootters inequality and entanglement \\
         sharing in discrete-variable systems}

The simplest conceivable quantum system in which
multipartite entanglement can arise is a system of three
two-level particles (qubits). Let two of these qubits,
say A and B, be in a maximally
entangled state (a Bell state). Then {\em no} entanglement is
possible between each of them and the third qubit C. In fact
entanglement between C and A (or B) would imply A and B being in a
mixed state, which is impossible because they are sharing a pure
Bell state. This simple observation embodies, in its sharpest
version, the {\em monogamy} of quantum entanglement\cite{monogamy},
as opposed to classical correlations which can be freely shared.

We find it instructive to look at this feature as a simple
consequence of the no-cloning theorem\cite{nocloning}. In fact, maximal couplewise
entanglement in both bipartitions AB and AC of a three-particle ABC system,
would enable perfect $1 \rightarrow 2$ telecloning\cite{telecloning} of an
unknown input state, which is impossible due to the linearity of
quantum mechanics. The monogamy constraints thus emerge as
fundamental properties enjoyed by quantum systems involving more than two
parties, and play a crucial role {\em e.g.}~in the security of
quantum key distribution schemes based on entanglement\cite{arturo},
limiting the possibilities of the malicious eavesdropper. Just like
in the context of cloning, where research is devoted to the problem
of creating the best possible {\em approximate} copies of a quantum
state, one can address the question of entanglement sharing in a
weaker form. If the two qubits A and B are still entangled but not in
a Bell state, one can then ask how much entanglement each of them
is allowed to share with qubit C, and what is the maximum genuine
tripartite entanglement that they may share all together.
The answer is beautifully encoded in the Coffman-Kundu-Wootters
(CKW) inequality\cite{CKW}
\begin{equation}\label{ckw3}
E^{A|(BC)} \, \ge \, E^{A|B \n
C} + E^{A|C \n B}\,,
\end{equation}
where $E^{A|(BC)}$ denotes the entanglement
between qubit $A$ and subsystem $(BC)$, globally in a state
$\varrho$, while $E^{A|B \n C}$ denotes
entanglement between $A$ and $B$ in the reduced state obtained
tracing out qubit $C$ (and similarly for
$E^{A|C \n B}$ exchanging the roles of $B$ and
$C$). \ineq{ckw3} states that the bipartite
entanglement between one single qubit,
say $A$, and all the others, is greater than the sum of all the possible
couplewise entanglements between $A$ and each other qubit.

\subsection{Which entanglement is shared?}

While originally derived for system of three qubits, it is natural,
due to the above considerations, to assume that \ineq{ckw3} be a
general feature of any three-party quantum system in
arbitrary (even infinite) dimensions. However, before proceeding,
the careful reader should raise an important question, namely {\em how}
are we measuring the bipartite entanglement in the different
bipartitions, and what the symbol $E$ stands for in \ineq{ckw3}.

Even if the system of three qubits is globally in a pure state,
its reductions will be obviously mixed. In fact, the various
physical processes responsible for the interpretation of the
{\em entropy of entanglement} as the unique proper entanglement
measure for pure states, cease to be equivalent for mixed states.
To give a typical example, one must spend more to produce a mixed
state $\varrho$ out of an ensemble of pure entangled states,
than what one earns back by distilling entanglement from $\varrho$
to a set of singlets via local operations and classical communication
(LOCC). Formally, the {\em entanglement of formation}\cite{eof}
$E_F$ is greater than the {\em distillable entanglement} $E_D$ of
generic mixed states, while both reduce to the entropy of
entanglement on pure states. More generally, there are several,
inequivalent measures of entanglement for mixed states, leading to
different {\em orderings} on the set of entangled states living in a
specified Hilbert space\cite{ordering}. Thus, a mixed
state $\varrho_A$ can be more entangled than another state
$\varrho_B$ with respect to a given measure, but less entangled than
$\varrho_B$ with respect to another measure.

In this piebald scenario (which we cannot further explore in this
paper), we should convince ourselves that different measures of
entanglement must be chosen, depending on the problem
one needs to address, and/or on the desired use of the entangled
resources. This picture is consistent, provided that each needed
measure is selected out of the cauldron of {\em
bona fide} entanglement measures, at least positive on
inseparable states and monotone under LOCC. Here we are addressing
the problem of entanglement sharing: one should not be so surprised to
discover that not all entanglement measures satisfy
\ineq{ckw3}. In particular, the entanglement of formation
fails to fulfill the task, and this fact led CKW to define, for
qubit systems, a new measure of bipartite entanglement
consistent with the quantitative monogamy constraint
expressed by \ineq{ckw3}.

\subsection{Entanglement of two qubits}

In discrete-variable systems, separability of a mixed state
$\varrho$ of two qubits (and of a system of one qubit and one
qutrit) is equivalent to the positivity of the partial
transpose\cite{PPT} (PPT) $\tilde \varrho$ of $\varrho$, defined as
the result of transposition performed on only one of the two
subsystems in some given basis. From a quantitative point of view, a
proper measure of entanglement is provided by the entanglement of
formation
\begin{equation}\label{entfor}
E_F(\varrho) \equiv \min_{\{p_i,\psi_i\}} \sum_i p_i E_v
(\ketbra{\psi_i}{\psi_i}) \, ,
\end{equation}
where the minimization is taken over those probabilities $\{p_i\}$
and pure states $\{\psi_i\}$ that realize the density matrix
$\varrho = \sum_i p_i \ketbra{\psi_i}{\psi_i}$, and $E_v$ is the
entropy of entanglement of $\ket{\psi_i}$. The latter is the von Neumann entropy
${\rm Tr}_{\rm A} \varrho_{\rm A}^i \log \varrho_{\rm A}^i$ of the
reduced density matrix $\varrho_{\rm A}^i = {\rm Tr}_{\rm B}
\ketbra{\psi_i}{\psi_i}$ obtained from the state $\ket{\psi_i}$ of
the bipartite system AB, by tracing out the degrees of freedom of
B. For two qubits, the entanglement of formation has been computed
by Wootters\cite{Wootters}, and reads
\begin{equation}
  E_F(\varrho) = f[\conc(\varrho)]\,, \label{eqwootters}
  \end{equation}
  with $f(x) = H [(1 + \sqrt{1-x^2})/2]$ and $H(x) = - x \log_2 x - (1-x) \log_2 (1-x)$.
 The quantity $\conc(\varrho)$ is called the
\emph{concurrence}\cite{Hill} of the state $\varrho$ and is defined
as $\conc(\varrho) \equiv
\max\{0,\sqrt{\lambda_1}-\sqrt{\lambda_2}-\sqrt{\lambda_3}-\sqrt{\lambda_4}\}\,
,$ where the $\{\lambda_i\}$'s are the eigenvalues of the matrix
$\varrho (\sigma_y \otimes \sigma_y) \varrho^{\ast} (\sigma_y
\otimes \sigma_y)$ in decreasing order, $\sigma_y$ is the Pauli spin
matrix and the star denotes complex conjugation in the computational
basis $\{\ket{ij}\equiv\ket{i}\otimes\ket{j},\;i,j=0,1\}$. Because
$f(x)$ is a monotonic convex function of $x \in [0,\,1]$, the
concurrence $\conc(\varrho)$ and its square, the
\emph{tangle}\cite{CKW} $\tau(\varrho)\equiv\conc^2(\varrho)$, are
proper entanglement monotones as well. On pure states, they are
monotonically increasing functions of the entropy of entanglement.
The concurrence coincides (for pure
states) with another entanglement monotone, the {\em
negativity}\cite{zircone}, defined in general as \be \label{nega}
{\cal N}(\varrho)=(\|\tilde \varrho \|_1-1)/2\: , \ee where
$\|\hat o\|_1=\,{\rm Tr}|\hat o|$ stands for the trace norm of the
hermitian operator $\hat o$. The quantity ${\cal N} (\varrho)$ is
equal to $|\sum_{i}\lambda_{i}|$, the modulus of the sum of the
negative eigenvalues of $\tilde\varrho$, quantifying the extent to
which the PPT criterion is violated. On the other hand, the tangle is
equal (for pure states $\ket\psi$) to the linear entropy of
entanglement $E_L$, defined as the linear entropy $S_L (\varrho_{\rm
A}) \equiv 1-{\rm Tr}_{\rm A}\varrho_{\rm A}^2$ of the reduced state
$\varrho_{\rm A} = {\rm Tr}_{\rm B} \ketbra{\psi}{\psi}$ of one
party.

\subsection{The residual tangle: a measure of tripartite
entanglement} \label{sec3tangle} After this survey, we can now
recall the crucial result that, for three qubits, the desired
measure $E$ such that the CKW inequality (\ref{ckw3}) is satisfied
is exactly the tangle\cite{CKW} $\tau$. The general definition of
the tangle, needed {\eg}to compute the leftmost term in \ineq{ckw3}
for mixed states, involves a convex roof analogous to that defined
in \eq{entfor}, namely
\begin{equation}\label{mixtangle}
\tau(\varrho) \equiv \min_{\{p_i,\psi_i\}} \sum_i p_i\,\tau
(\ketbra{\psi_i}{\psi_i})\,.
\end{equation}
With this general definition, which implies that the tangle is a
convex measure on the set of density matrices, it was sufficient for
CKW to prove \ineq{ckw3} only for pure states of three qubits, to
have it satisfied {\em for free} by mixed states as well\cite{CKW}.

Once one has established a monogamy inequality like \ineq{ckw3}, the
following natural step is to study the difference between the
leftmost quantity and the rightmost one, and to
interpret this difference as the {\em residual entanglement},
not stored in couplewise correlations, that thus quantifies
the genuine tripartite entanglement shared by the three qubits.
The emerging measure
\begin{equation}\label{tau3}
\tau_3^{A|B|C} \equiv \tau^{A|(BC)} - \tau^{A|B \n
C} - \tau^{A|C \n B}\,,
\end{equation}
known as the {\em three-way tangle}\cite{CKW}, has indeed some nice
features. For pure states, it is invariant
under permutations of any two qubits, and more remarkably it has been
proven to be a tripartite entanglement monotone under
LOCC\cite{wstates}. However, no operational interpretation for the
three-tangle, possibly relating it to the optimal distillation rate of
some canonical `multiparty singlet', is currently known. The
reason lies probably in the fact that the notion of a well-defined
maximally entangled state becomes fuzzier when one moves to the
multipartite setting. In this context, it has been shown
that there exist two classes of three-party fully inseparable
pure states of
three qubits, inequivalent under stochastic LOCC operations, namely
the Greenberger-Horne-Zeilinger (GHZ) state\cite{ghzs}
$\ket{\psi_{\rm GHZ}} = (1/\sqrt{2}) \left[\ket{000} +
\ket{111}\right]$, and the $W$ state\cite{wstates} $\ket{\psi_{W}} =
(1/\sqrt{3}) \left[\ket{001} + \ket{010} + \ket{100}\right]$. From the
point of view of entanglement, the
big difference between them is that the GHZ state has maximum
residual three-party tangle [$\tau_3(\psi_{\rm GHZ}) = 1$] with zero
couplewise quantum correlations in any two-qubit reduction, while
the $W$ state contains maximum two-party entanglement between any
couple of qubits in the reduced states and it consequently saturates
\ineq{ckw3} [$\tau_3(\psi_W) = 0$].

\subsection{Monogamy inequality for $N$ parties}

So far we have recalled the known
results on the problem of entanglement sharing in systems of three
parties, leading to the definition of the residual tangle as a
proper measure of genuine tripartite entanglement for three qubits.
However, if the monogamy of entanglement is really a universal
property of quantum systems, one should aim at finding more general
results.

There are two axes along which one can move, pictorially, in this respect.
One direction concerns the investigation on distributed entanglement
in systems of more than three parties, starting with the simplest
case of $N \ge 4$ qubits (thus moving along the horizontal axis of increasing number
of parties). On the other hand, one should analyze the sharing
structure of multipartite entanglement in higher dimensional
systems, like qudits, moving, in the end, towards continuous variable (CV)
systems
(thus going along the vertical axis of increasing Hilbert space
dimensions). The final goal would be to cover the entire square
spanned by these two axes, in order to establish a really complete
theory of entanglement sharing.

Let us start moving to the right. It is quite natural to expect
that, in a $N$-party system, the entanglement between qubit $p_i$
and the rest should be greater than the total two-party entanglement
between qubit $p_i$ and each of the other $N-1$ qubits. So the
generalized version of \ineq{ckw3} reads
\begin{equation}\label{ckwn}
E^{p_i|{\cal P}_i} \, \ge \, \sum_{j \ne i} E^{p_i|p_j}\,,
\end{equation}
with ${\cal P}_i \equiv (p_1,\ldots,p_{i-1},p_{i+1},\ldots,p_N)$.
Proving \ineq{ckwn} for {\em any} quantum system in arbitrary
dimension, would definitely fill the square; it appears though as a
formidable task. However, partial, encouraging results have been
recently obtained.

Osborne and Verstraete have shown that the generalized monogamy
inequality (\ref{ckwn}) holds true for any (pure or mixed) state of
a system of $N$ qubits\cite{osborne}, proving a longstanding
conjecture due to CKW themselves\cite{CKW}. Again, the entanglement
has to be measured by the tangle $\tau$. This is an important
result; nevertheless, one must admit that, if more than three
parties are concerned, it is not so obvious why all the bipartite
entanglements should be decomposed only with respect to a single
elementary subsystem.
One has in fact an exponentially increasing number of ways to
arrange blocks of subsystems and to construct multiple splittings of
the whole set of parties, across which the bipartite (or, even more
intriguingly, the multipartite) entanglements can be compared. This
may be viewed as a third, multifolded axis in our `geometrical'
description of the possible generalizations of \ineq{ckw3}.
Leaving aside in the present paper this intricated plethora of
additional situations, we stick to the monogamy constraint of
\ineq{ckwn}, obtained decomposing the bipartite entanglements with
respect to a single particle, while keeping in mind that for more
than three particles the residual entanglement emerging from
\ineq{ckwn} is not necessarily {\em the} measure of multipartite
entanglement. Rather, it properly quantifies the  entanglement not
stored in couplewise correlations, and thus finds interesting
applications for instance in the study of quantum phase transitions
and criticality in spin systems\cite{quph,verrucchi}.

\subsection{Entanglement sharing among qudits}

The first problem one is faced with when trying to investigate the
sharing of quantum correlations in higher dimensional systems is to
find the correct measure for the quantification of bipartite
entanglement. Several approaches to generalize Wootters' concurrence
and/or tangle have been developed\cite{concext}. In the present
context, maybe the most relevant result has been recently obtained
by Yu and Song\cite{cinesi}, who established the general monogamy
inequality (\ref{ckwn}) for an arbitrary number of qudits
({\ie}$d$-dimensional quantum systems), for any finite $d$. They
used a generalization of the tangle $\tau$, defined for
mixed states as the convex-roof extension \eq{mixtangle} of the
linear entropy of entanglement $E_L$ for pure states. Moreover, the
authors claim that the corresponding residual tangle is a
proper measure of multipartite entanglement. Let us remark however
that, at the present stage in the theory of entanglement sharing,
trying to make sense of a heavy mathematical framework (within
which, moreover, a proof of monotonicity of the $N$-way tangle under
LOCC has not been established yet for $N >3$, not even for qubits)
with little, if any, physical insight, is likely not worth trying.
Probably the CKW inequality is interesting not because of the
multipartite measure it implies, but because it embodies a
quantifiable trade-off between the distribution of bipartite
entanglement.

In this respect, it seems relevant to address the following
question, raised by Dennison and Wootters\cite{qudits}. One is
interested in computing the maximum possible bipartite entanglement
between {\em any} couple of parties, in a system of three or more
qudits, and in comparing it with the entanglement capacity $\log_2
d$ of the system. Their ratio $\varepsilon$ would provide an
immediate quantitative bound on the shareable entanglement, stored
in couplewise correlations. Results obtained for $d=2$, $3$
and $7$ (using the entanglement of formation) suggest for three
qudits a general trend of increasing $\varepsilon$ with increasing
$d$\cite{qudits}. While this is only a preliminary analysis, it
raises intriguing questions, pushing the interest in entanglement
sharing towards infinite-dimensional systems. In fact,
if $\varepsilon$ saturated to $1$ for $d \rightarrow \infty$, this
would entail the really counterintuitive result that entanglement
could be freely shared in this limit! We notice that, being the
entanglement capacity infinite for $d \rightarrow \infty$,
$\varepsilon$ vanishes if the maximum  couplewise entanglement is
not infinite. And this is the case, because again an infinite shared
entanglement between two two-party reductions would allow perfect $1
\rightarrow 2$ telecloning exploiting Einstein-Podolski-Rosen (EPR)\cite{epr}
correlations, but this is forbidden by quantum mechanics.
Nevertheless, the study of entanglement sharing in CV
systems yields surprising consequences.

\section{Entanglement sharing in continuous variable systems}

The first study of entanglement sharing in CV
systems has been performed in Ref.~[\refcite{contangle}], focusing on
the physically relevant class of Gaussian states.

\subsection{Entanglement of Gaussian states}
In a CV system consisting of $N$ canonical modes, associated to an
infinite-dimensional Hilbert space, and described by the vector
$\hat{X}$ of the field quadrature operators, Gaussian states (such
as squeezed, coherent and thermal states) are those states
characterized by first and second moments of the canonical
operators. When addressing physical properties invariant under local
unitary operations, like entanglement, first moments can be
neglected and Gaussian states can then be fully described by the $2N
\times 2N$ real covariance matrix (CM) $\gr{\sigma}$, whose entries
are $\sigma_{ij}=1/2\langle\{\hat{X}_i,\hat{X}_j\}\rangle$. A
physical CM $\gr{\sigma}$ must fulfill the uncertainty relation
$\gr{\sigma}+i\Omega \geq 0$, with the symplectic form
$\Omega=\oplus_{i=1}^{n}\omega$ and $\omega=\delta_{ij-1}-
\delta_{ij+1},\, i,j=1,2$. In phase space, any $N$-mode Gaussian
state can be written as $\gr{\sigma}= S^T \gr{\nu} S$, with
$\gr{\nu}=\,{\rm diag}\,\{n_1,n_1,n_2,n_2, \ldots, n_N, n_N \}$ and
$S$ a symplectic operation. The set $\Sigma=\{n_i\}$ constitutes the
symplectic spectrum of $\gr{\sigma}$ and its elements must fulfill
the conditions $n_i\ge 1$, ensuring positivity of the density matrix
$\varrho$ associated to $\gr{\sigma}$. The degree of purity
$\mu=\,{\rm Tr}\,\varrho^2$ of a Gaussian state with CM
$\gr{\sigma}$ is simply $\mu=1/\sqrt{\,{\rm Det}\,\gr{\sigma}}$.

Concerning the entanglement, the PPT criterion is again a necessary and
sufficient condition for separability of $(N+1)$-mode Gaussian
states of $(1 \times N)$-mode bipartitions\cite{simon00,werner02}
and of $(M+N)$-mode bisymmetric Gaussian states of $(M \times
N)$-mode bipartitions\cite{unitarily}. In phase space, partial
transposition with respect to a $(1 \times N)$-mode bipartition
amounts to a mirror reflection of one quadrature associated to the
single-mode party. If $\{\tilde{n}_i\}$ is the symplectic spectrum
of the partially transposed CM $\tilde{\gr{\sigma}}$, then a
$(N+1)$-mode Gaussian state with CM $\gr{\sigma}$ is separable if
and only if $\tilde{n}_i\ge 1$ $\forall\, i$. This implies that {\em
bona fide} measures of CV entanglement are the negativity $\N$
\eq{nega} and, more properly, the {\em logarithmic
negativity}\cite{vidal02}
\begin{equation}\label{logneg}
E_{\N} \equiv \log \|\tilde \varrho \|_1\,,
\end{equation}
which is readily computed in terms of the symplectic spectrum
$\tilde{n}_i$ of $\tilde{\gr{\sigma}}$ as
$E_{\N}=-\sum_{i:\tilde{n}_i<1}\log \tilde{n}_i$. The logarithmic
negativity is additive on tensor product states and constitutes an
upper bound on the distillable entanglement. For two-mode symmetric
Gaussian states only, the entanglement of formation \eq{entfor} has
been computed\cite{efprl} and it is completely equivalent to $E_\N$
in that subcase.

\subsection{The continuous variable tangle}

After this brief introduction on Gaussian states (see Ref.
[\refcite{brarev}] for a recent review), let us now look for the
proper measure of bipartite entanglement, which would be the CV
analogue of the tangle. While a formal, mathematically justified
definition of this new measure has been given in Ref.
[\refcite{contangle}], we believe it is more instructive to follow,
here, a simple trial and error strategy to arrive at the correct
result.

We can reasonably assume that \ineq{ckw3} is true for three-mode
Gaussian states, like it should be for any three-party quantum
system. The problem is to find the proper measure $E$. The first
attempt is naturally to use the entanglement of formation (when
computable) or the negativities. Immediate inspection reveals that
they actually fail, even in the simplest instance of pure, fully
symmetric, three-mode Gaussian states\cite{contangle}. So, next
trial. Let us construct a generalization of the tangle via the
convex roof like in \eq{mixtangle}, where for pure states the tangle
is defined as the linear entropy of entanglement, just like in the
case of qubits and qudits. The corresponding tangle for CV systems would range from $0$
to $1$, which is uncommon when dealing with states whose entanglement can
be infinite; and, in fact, this candidate, which works fine for
the quantification of entanglement sharing in any finite dimension
$2 \le d <\infty$, fails for $d=\infty$, for instance on
pure, bisymmetric\cite{unitarily} three-mode Gaussian states.
This could {\em prima facie} discourage us from trying to define a CV tangle; indeed, there is
another chance left, thanks to the following crucial observation.
For mixed states of two qubits, the tangle can
be viewed {\em equivalently} as the convex-roof extension of the squared negativity
(the latter coinciding with the concurrence for pure states).
This fact then suggests to define a CV tangle via the negativity or, better, via the
logarithmic negativity. In fact, if the monogamy inequality is
satisfied using a measure $E$ of bipartite entanglement, it will
hold as well using any other increasing and convex
function of $E$. This is exactly the case for negativities,
because $\N$ is a convex function of $E_\N$.

From the above considerations, it follows that a privileged
candidate to comply with the CV versions of the monogamy inequalities
(\ref{ckw3},\ref{ckwn}) is thus the convex-roof extension of the squared
logarithmic negativity, which will define the continuous-variable
tangle, or, in short, the {\em contangle}\cite{contangle} $E_\tau$.
For a generic pure state $\ket{\psi}$ of a $(1+N)$-mode CV system,
we can formally define the contangle as
\begin{equation}\label{etaupure}
E_\tau (\psi) \equiv \log^2 \| \tilde \varrho \|_1\,,\quad \varrho =
\ketbra\psi\psi\,.
\end{equation}
$E_\tau (\psi)$ is a proper measure of bipartite entanglement, being
a convex,
 increasing function of the logarithmic negativity
$E_\N$, which is equivalent to the entropy of entanglement for
arbitrary pure states. In the case of a pure Gaussian state
$\ket\psi$ with CM $\sig^p$, $E_\tau (\sig^p) = \log^2 (1/\mu_1 -
\sqrt{1/\mu_1^2-1})$, where $\mu_1 = 1/\sqrt{\det\sig_1}$ is the
local purity of the reduced state of mode $1$, described by a CM
$\sig_1$ (we are dealing with a $1 \times N$ bipartition).
Definition (\ref{etaupure}) is naturally extended to generic mixed
states $\varrho$ of $(N+1)$-mode CV systems through the convex-roof
formalism, namely
\begin{equation}\label{etaumix}
E_\tau(\varrho) \equiv \inf_{\{p_i,\psi_i\}} \sum_i p_i
E_\tau(\psi_i)\,,
\end{equation}
where the infimum is taken over the decompositions of $\varrho$ in
terms of pure states $\{\ket{\psi_i}\}$.  Dealing with infinite
Hilbert spaces the index $i$ is continuous, so the sum in
\eq{etaumix} should be replaced by an integral, and the
probabilities $\{p_i\}$ by a distribution $\pi(\psi)$. Let us recall
that any multimode mixed Gaussian state with CM $\sig$, admits a
decomposition in terms of an ensemble of pure Gaussian states. The
infimum of the average contangle, taken over all pure Gaussian
decompositions only, defines the {\em Gaussian contangle}
$G_{\tau}$, which is an upper bound to the true contangle $E_\tau$,
and an entanglement monotone under Gaussian local operations and
classical communications (GLOCC)\cite{geof}. The Gaussian contangle,
similarly to the Gaussian entanglement of formation\cite{geof},
acquires the simple form $G_\tau (\sig) \equiv \inf_{\sig^p \le
\sig} E_\tau(\sig^p)$, where the infimum runs over all pure Gaussian
states  with CM $\sig^p \le \sig$.

Equipped with these properties and definitions, one can prove
a series of results\cite{contangle}. In particular,
\ineq{ckwn} is satisfied by all pure three-mode and all pure
{\em symmetric} $N$-mode Gaussian states, using either
$E_\tau$ or $G_\tau$ to quantify bipartite entanglement, and by
all the corresponding mixed states using $G_\tau$. Furthermore,
there is numerical
evidence supporting the conjecture that the general CKW \ineq{ckwn}
should hold for all {\it nonsymmetric} $N$-mode Gaussian states as well.

The sharing constraint (\ref{ckw3}) leads to the definition of the
{\em residual contangle} as a tripartite entanglement quantifier.
However, for generic three-mode Gaussian states the residual contangle is
partition-dependent. In this respect, a proper
quantification of tripartite entanglement is provided by the
{\em minimum} residual contangle
\begin{equation}\label{etaumin}
E_\tau^{i|j|k}\equiv\min_{(i,j,k)} \left[
E_\tau^{i|(jk)}-E_\tau^{i|j}-E_\tau^{i|k}\right]\,,
\end{equation}
where $(i,j,k)$ denotes all the permutations of the three mode
indexes. This definition ensures that $E_\tau^{i|j|k}$ is invariant
under mode permutations and is thus a genuine three-way property of
any three-mode Gaussian state. We can adopt an analogous definition
for the minimum residual Gaussian contangle $G_\tau^{i|j|k}$. One
finds that the latter is a proper measure of genuine tripartite
CV entanglement, because it can be proven to be
an entanglement monotone under tripartite GLOCC
for pure three-mode Gaussian states\cite{contangle}.

\subsection{Promiscuous sharing of continuous variable entanglement}

Let us now analyze the sharing structure of CV entanglement by
taking the residual contangle as a measure of tripartite
entanglement, in analogy with the study done for three
qubits\cite{wstates}. Namely, we pose the problem of identifying the
three-mode analogues of the two fully inseparable and symmetric three-qubit
pure states, the GHZ state\cite{ghzs} and the $W$ state\cite{wstates},
discussed in Sec.~\ref{sec3tangle}. Surprisingly enough, in
symmetric three-mode Gaussian states, if one aims at maximizing (at
given single-mode squeezing) either the two-mode contangle
$E_\tau^{i|l}$ in any reduced state ({\it i.e.}~aiming at the CV
$W$-like state), or the genuine tripartite contangle ({\it
i.e.}~aiming at the CV GHZ-like state), one finds the same, unique
family of pure symmetric three-mode squeezed states. These states,
previously named ``GHZ-type'' states\cite{brarev}, can be defined
for generic $N$-mode systems, and their multimode entanglement
scaling can be studied\cite{adescaling,unitarily}. The peculiar
nature of entanglement sharing in this class of CV GHZ/$W$ states is
further confirmed noting that if one requires $E_\tau^{i|(jk)}$ to
be maximum under the constraint of separability of all two-mode
reductions, one finds states whose residual contangle is strictly
smaller than the one of the GHZ/$W$ states, at fixed squeezing.

Therefore, in symmetric three-mode Gaussian states, when there is no
two-mode entanglement, the three-party one is not enhanced, but
frustrated.
These results, unveiling a major difference between
discrete-variable and CV systems, establish the {\em promiscuous}
nature of entanglement sharing in symmetric Gaussian states. Being
associated with degrees of freedom with continuous spectra, states
of CV systems need not saturate the sharing inequality to achieve
maximum couplewise correlations. In fact, without violating the
monogamy constraint (\ref{ckw3}), pure symmetric three-mode Gaussian
states are maximally three-way entangled and, at the same time,
maximally robust against the loss of one of the modes due, for
instance, to decoherence.

Finally, the residual contangle \eq{etaumin} in this class of
GHZ/$W$ states acquires a clear operative meaning in terms of the
optimal fidelity in a three-party CV teleportation
network\cite{telepoppate}. This result enforces the
interpretation of the contangle $E_\tau$ as a {\em bona fide}
measure of tripartite entanglement for Gaussian states,
because it appears as the most natural infinite-dimensional
extension of the tangle $\tau$ which quantifies entanglement
sharing among qubits and qudits.

\section*{Acknowledgments}

Financial support from MIUR, INFN, and INFM is acknowledged. We are
grateful to W. K. Wootters for lots of suggestions and enjoying
discussions on the subject.

\paragraph{Note added in proof.} Recently, Hiroshima, Adesso and Illuminati
have proven inequality (\ref{ckwn}) for all Gaussian states of
$N$-mode CV systems\cite{40}, by using the Gaussian tangle defined
as the (convex-roof extended) squared negativity. This fundamental
result extends the findings of Ref. \refcite{contangle} and
establishes the general monogamy of distributed Gaussian
entanglement. The conjecture raised in Sec. 2.2 is indeed true.

\end{document}